\documentclass[twocolumn, secnumarabic, amssymb, noshowpacs, aps, showpacs,prb]{revtex4-1}
\usepackage{graphics}
\usepackage[dvips]{graphicx}
\usepackage{dcolumn}
\usepackage{amsmath}
\usepackage{lipsum}

\begin{document}
\title{Why the effective-mass approximation works so well for nano-structures}
\author{Pedro Pereyra}
\address{F\'{i}sica Te\'{o}rica y Materia Condensada, UAM-Azcapotzalco, C.P. 02200, Ciudad de M\'{e}xico, M\'{e}xico }
\date{\today}

\begin{abstract}
The reason why the effective-mass approximation works so well with nanoscopic structures has been an enigma and a challenge for theorists. To explain this issue, we re-derive the effective-mass approximation using, instead of the wave functions for {\it infinite}-periodic-systems and the ensuing continuous bands, the eigenfunctions and eigenvalues obtained in the theory of finite periodic systems, where the {\it finiteness} of the number of primitive cells in the nanoscopic layers, is a prerequisite and an essential condition. This derivation justifies and shows why this approximation works so well for nano-structures. We show also with explicit optical-response calculations that the rapidly varying eigenfunctions $\Phi_{\epsilon_0,\eta_0}(z)$ of the one-band wave functions $\Psi^{\epsilon_0,\eta_0}_{\mu,\nu}(z)= {\it \Psi}^{\epsilon_0}_{\mu,\nu}(z) \Phi_{\epsilon_0,\eta_0}(z)$, can be safely dropped out for the calculation of inter-band transition matrix elements.
\end{abstract}

\maketitle

\section{Introduction}
The effective-mass approximation (EMA) is, without a doubt, the most recurrent and widely used approximation in theoretical calculations involving semiconductor structures. The formal justification of why this approximation, where the wave packets are constructed in terms of {\it infinite} periodic system wave functions,\cite{Wannier,Slater,Luttinger,Altarelli,AltarelliLesHuches,Pollak,Dingle} works so well for {\it finite} micro and nano-structures, has been an enigma and a challenge for theorists. Despite the various guises of the EMA, the correct explanation has remained elusive.  M. G. Burt in a number of papers\cite{Burt} analysed critically the drawbacks of the ``conventional" EMA,  and tried to overcome these attempts providing a ``new" envelope-function method, using again wave functions of {\it infinite} periodic systems. Now that the theory of finite periodic systems (TFPS) has evolved and has shown the ability to obtain the true, {\it bona fide}, energy eigenvalues and eigenfunctions of finite periodic structures with a finite number of unit cells,\cite{Abeles,Erdos,Claro1982,Ricco,Vezzetti,Kolatas,Griffiths,Peisakovich,
PereyraPRL,PereyraJPA,PereyraCastillo} it is worth reviewing and re-deriving the EMA within the TFPS  to understand why it works so well. The purpose of this letter is to re-derive the effective mass approximation taking into account the system and layers finiteness as the fundamental requisite.

Superlattices and layered structures are characterized by the simultaneous presence of two length scales: the crystalline unit cells in the semiconductor layers of atomic size and the layers widths. While the primitive cells lengths are of the order of 0.5nm, depending on the the semiconductor, the layers widths are of the order of 5nm, depending on the number of atomic cells per layer. This important difference in size is behind the factorization of the heterostructure wave function (HWF) in terms of rapid and slowly varying functions. The finiteness of the number $n_X$ of primitive cells, in the direction of growth, of layer $X$ (=A,B,...), and the finiteness of the number of layers in the heterostructure or number of superlattice (SL) unit cells $n_S$, is not only an obvious characteristic, but also an essential requisite in the TFPS.

\section{Finiteness of periodic layers. An outline of the TFPS}

Soon after the semiconductor
SLs were introduced,\cite{Keldysh1962,EsakiTsu1970} and the subbands (or minibands)
structures of direct and indirect band gap semiconductors were
experimentally and theoretically
confirmed,\cite{Esaki1972,Dingle1974,Mukherji1975,Miller1976,Chang1977,
SaiHalaszChang1978,Capasso1986,LuoFurdyna1990,
Rauch1997,Petrov1997,Heer1998} Leo Esaki noticed that whereas in
reality SLs contain a finite number of layers, with a finite
number of atomic cells each, the standard  theoretical approaches
tacitly assume that SLs  are {\it infinite}-periodic structures with
alternating layers containing also an  {\it infinite} number of atomic
cells.\cite{EsakiLesHuches} In fact, the HWF and SL wave functions are generally\cite{Luttinger,Altarelli,Dresselhauss,Breitenecker,Sanders,Bastard1987NATO,
Smith,Baraff} written
as $\psi ({\bf r})=\sum_l u_{n_{l}}({\bf r})f_{l}({\bf r})$,  with
$u_{n_{ l}}({\bf r})$ the periodic part of the host-semiconductor
Bloch's function at band $n_{l}$, and $f_{l}({\bf r})\propto\exp[{i
{\bf k}_{\perp}\cdot {\bf r}_{\perp}}]\chi_{l}(z)$ the envelope wave
function, with ${\bf k}_{\perp}= {\bf k}_x+{\bf k}_y$ the perpendicular wave number assumed, generally, a constant of motion.\cite{Bastard1987NATO} At the end, it is common to assume wave functions $\psi({\bf r})$ set up from wave functions $u_{n_{0}}$ of only one band, evaluated at the center of the Brillouin zone
or at the subband edge ${\bf k} = 0$. For SLs the envelope function is, again, written in terms of Bloch-type
functions $\chi_{\mu}(z)$ $=\exp(iqz)u_{\mu}(z)$, characterized by
a subband index $\mu$ and a continuous wave number $q$ that is
then artificially discretized, via the cyclic boundary condition.

On the other side, the theory of finite periodic systems has grown,
and has been generalized to include periodic structures with arbitrary potential profiles, arbitrary but finite number $n$ of unit cells
and arbitrary but finite number $N$ of propagating modes for open, bounded and
quasi-bounded periodic structures.\cite{PereyraPRL,PereyraJPA,PereyraCastillo,Pereyra2005}
The TFPS is based on the transfer matrix properties and the rigorous fulfillment of continuity conditions, that make possible to express the $n$-cells transfer matrix $M_n$ as $M^n$, where $M$, for time reversal invariant systems, is the single-cell transfer matrix of dimention $2N$$      \times$$2N$
\begin{eqnarray}
M(z_{i+1},z_{i})=\left( \begin{array}{cc} \alpha & \beta \cr \beta^* & \alpha^*  \end{array}\right).
\end{eqnarray}
The accurate calculation of this matrix is crucial in this approach. The complex matrix functions $\alpha$ and $\beta$ depend strongly on the atomic or heterostructure potential profiles. The relation
\begin{eqnarray} M_n=M^n=\left( \begin{array}{cc} \alpha_n & \beta_n \cr \beta_n^* & \alpha_n^*  \end{array}\right),
\end{eqnarray}
that was the source of errors in numerical calculations,\cite{Luque} has been rigorously transformed, after defining the matrix function $p_{n-1}=\beta^{-1}\beta_n$, into the matrix-recurrence relation\cite{PereyraPRL,PereyraJPA}
\begin{equation}\label{RecRelation}
p_n-(\beta^{-1} \alpha \beta+ \alpha^*)p_{n-1}+p_{n-2}=0,
\end{equation}
with analytic solutions. In the single mode approximation, of interest here, this relation becomes the recurrence relation of Chebyshev polynomials of the second kind $U_n$, evaluated at the real part of $\alpha=\alpha_R+i\alpha_I$. The $n$-cell transfer matrix elements, $\alpha_n$ and $\beta_n$, can straightforwardly be determined, through the simple relations
\begin{eqnarray}
\alpha_n=U_n-\alpha^*U_{n-1},  \hspace{0.2in} {\rm and} \hspace{0.2in}\beta_{n}=\beta U_{n-1}.
\end{eqnarray}
The eigenvalues of any quasi-bounded (qb) periodic system defined between $z_L$ and $z_R$, see figure 1, with $z_0-z_L=z_R-z_n=d/2$,  can be obtained by solving the equation\cite{Pereyra2005}
\begin{eqnarray}\label{EqEigenv}
\!\!{\rm Re}\left(\alpha_ne^{ikd}\right)\!-\!\frac{k^{2}\!-\!q_{w}^{2}}{2q_{w}k}{\rm Im}\left(\alpha_ne^{ikd}\right)\!-\!\frac{k^{2}\!+\!q_{w}^{2}}{2q_{w}k}\beta_{nI}\!=\!0.
\end{eqnarray}
$q_w$ and $k$ are the wave numbers at the left (right) and right (left) of the discontinuity point $z_L$ ($z_R$) and $\beta_{nI}$ the imaginary part of $\beta_n$.
The eigenfunctions of the quasi-bounded superlattice are given by\cite{Pereyra2005}
\begin{eqnarray}\label{EquEigFunc1}
{\it \Psi}_{\mu,\nu }^{qb}\!&&(z)=\!\frac{a_o e^{\!i k d/2}}{2k}\Bigl[\Bigl((\alpha_{p}\!+\!\gamma
_{p})\alpha
_{j}\!+\!(\beta_{p}\!+\!\delta_{p})\beta_{j}^{\ast}\Bigr)(k\!-\!iq_{w}
)\Bigr. \nonumber \\ &\!\!+\!\!&\Bigl.\Bigl((\alpha
_{p}\!+\!\gamma_{p})\beta_{j}\!+\!(\beta_{p}\!+\!\delta _{p})\alpha _{j}^{\ast
}\Bigr) e^{\!-i k d}(k\!+\!iq_{w})\Bigr]_{E=E_{\mu,\nu}},\nonumber \\
\end{eqnarray}
with $a_o$ a normalization constant
and $z$ any point in the $j+1$ cell. $\alpha_j$,
$\beta_j$,... are matrix elements of the transfer matrix $M_j(z_j,z_0)$ that connects the state vectors  at points separated by exactly $j$ unit cells. $\alpha_p$,
$\beta_p$ ... , where $p$ stands for part of a unit cell, are the matrix elements of the transfer matrix $M_p(z,z_j)$ that connects the state vectors at $z_j$ and $z$, for $z_j\leq z \leq z_{j+1}$.

Our purpose here is to derive the effective mass approximation for the Schr\"odinger equation of a layered semiconductor heterostructure $A/B/C...$, using the eigenvalues and  eigenfunctions obtained in the TFPS. We will assume, without loss of generality, that our system is a binary structure $A/B/A...B/A$, where the periodic semiconductor layers $A=(a_A)^{n_A}$ and $B=(b_B)^{n_B}$ contain $n_A$ and $n_B$ unit cells $a_A$ and $b_B$, respectively, in the growing direction $z$. We will show that the effective-mass approximation (EMA) can be derived when the heterostructure wave function $\psi(z)$ is written as the product $\Phi_{\epsilon_{0},\kappa_{0}}(z)$ ${\it \Psi}^{\epsilon_0}_{\mu,\nu}(z)$, where ${\it \Psi}^{\epsilon_0}_{\mu,\nu}(z)$ is the envelope function and $\Phi_{\epsilon_{0},\kappa_{0}}(z)$ is the fast-varying function obtained in the TFPS, evaluated at the band-edges defined by the energy band index $\epsilon_{0}$ and the intra-band (or wave number) index $\kappa_{0}$. In the particular case of periodic heterostructures, i.e. of SLs $(AB)^n=((a_A)^{n_A}(b_B)^{n_B})^n$, the envelope functions are straightforwardly obtained in the EMA and the TFPS. It is worth emphasizing that since the transfer matrices are the matrix representation of the continuity and boundary conditions and the phase evolution of the quantum states, it is clear that the fast-varying and envelope wave functions, obtained in the TFPS, fulfill the continuity and boundary conditions. We will show also, for a specific example, that the optical response calculated with the matrix elements $\langle {\it \Psi}^{\epsilon'_0}_{\mu',\nu'} \Phi^{A}_{\epsilon',\kappa'}(z)|H_{\rm int}|{\it \Psi}^{\epsilon_0}_{\mu,\nu} \Phi^{A}_{\epsilon,\kappa}(z)\rangle$ is practically the same as the optical response obtained with the matrix elements $\langle {\it \Psi}^{\epsilon'_0}_{\mu',\nu'}|H_{\rm int}|{\it \Psi}^{\epsilon_0}_{\mu,\nu} \rangle$, were the fast-varying wave functions $\Phi^{A}_{\epsilon,\kappa}(z)$ are ignored.

\begin{figure}
\begin{center}
\includegraphics [width=240pt]{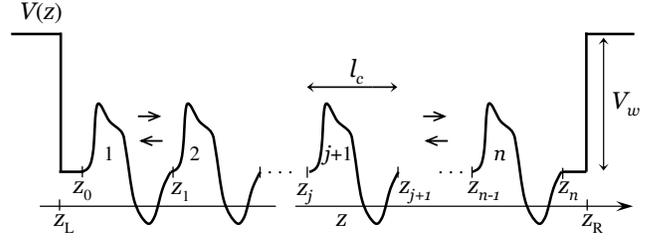}
\caption{Parameters of a quasi-bounded superlattice whose unit-cell has an arbitrary potential shape. The wave
function in Eq. (\ref{EquEigFunc1}) is defined at any point $z$ of the $j\!+\!1$
cell, with $0\leq j \leq (n-1)$.} \label{f1}
\end{center}
\end{figure}

\section{An alternative derivation of the effective-mass approximation}
\begin{figure*}
\begin{center}
\includegraphics*[width=340pt]{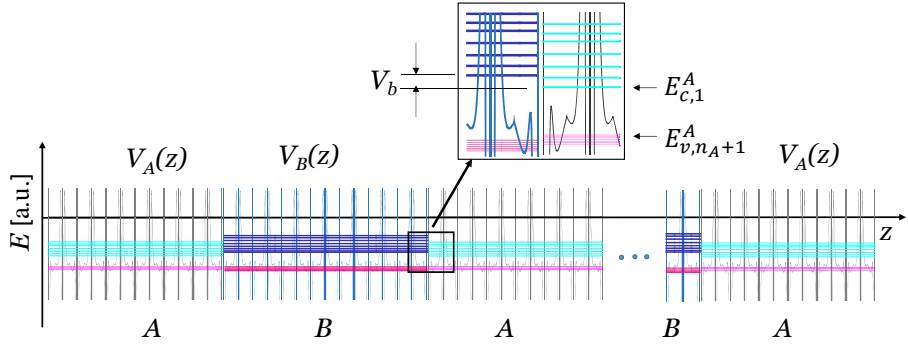}
\caption{The energy bands and the rapidly varying periodic potentials $V_A (z)$ and $V_B (z)$ in the layered semiconductor structure $A/B/A...B/A$, with $n_A$=6 and $n_B$=7, respectively. In the inset, details of the atomic potentials and band structures at layers interface when $A/B$ is, say, GaAs/AlGaAs. The band-edge energies $E^A_{v,n_A+1}$ and $E^A_{c,1}$ and the band split off $V_b$ are shown also.} \label{f2}
\end{center}
\end{figure*}
\begin{figure*}
\begin{center}
\includegraphics*[width=340pt]{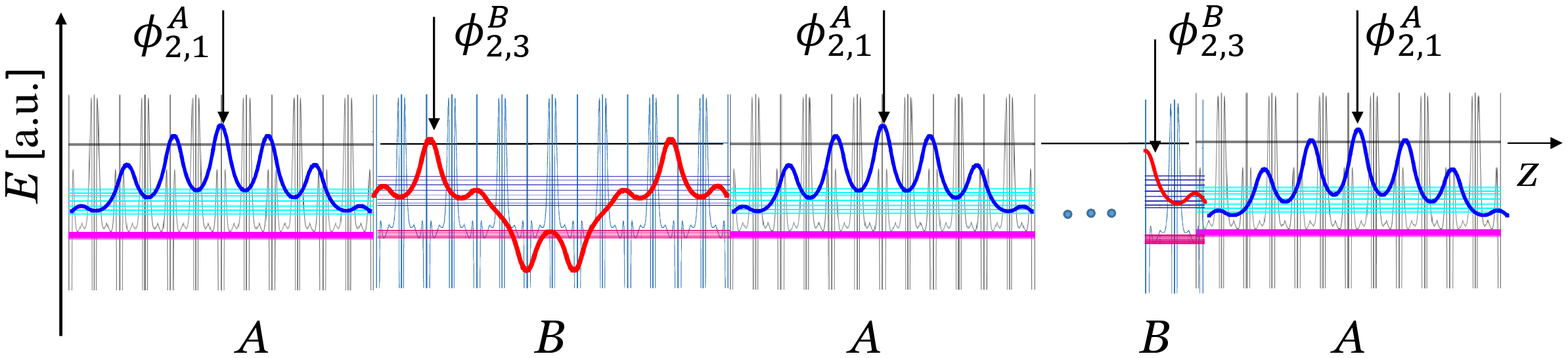}
\caption{The heterostructure wave function $\Phi^{\epsilon_B,\kappa_B}_{\epsilon_A,\kappa_A}(z)$, defined in Equation (\ref{SLWaveFunction}), for $\epsilon_A=\epsilon_B=2$, $\eta_A=1$ and $\eta_B=3$.} \label{f3}
\end{center}
\end{figure*}
Suppose now that for each layer $X$ (with $X$ equal $A$ or $B$) we can write the one-particle Schr\"odinger equation
\begin{eqnarray}\label{Hamiltonian}
\left( \frac{{\bf p}^2}{2m} + V_X({\bf r}) \right) \Phi^X({\bf r})=E\Phi^X({\bf r}),\hspace{0.3in}
\end{eqnarray}
where the potential $V_X({\bf r})$ is periodic, at least in the growing direction $z$. To simplify this problem we can follow the confined geometry method in Ref. [\onlinecite{Bagwell}] and the multichannel transfer matrix method in Refs. [\onlinecite{PereyraPRL}] and [\onlinecite{PereyraJPA}]. If we assume that the transverse widths are $w_x$ and $w_y$ and we write the potential $V_X({\bf r})$ as the sum of a confining potential $V_{X}^C(x,y)$, which is infinite for $|x|>w_x/2$ and $|y|>w_y/2$, and the function $V_{X}^L(x,y,z)$ periodic in $z$, the orthonormal wave functions $\chi_j(x,y)$, which are solutions of
\begin{eqnarray}\label{ConfiningEq}
\left(\!-\frac{\hbar^2}{2m}\Bigl(\frac{\partial^2}{\partial x^2}\!+\!\frac{\partial^2}{\partial y^2}\!\Bigr)\!+\!V_{X}^C(x,y)\!\right)\! \chi_j^X(x,y)\!=\!\varepsilon_j^X\chi_j^X(x,y),\hspace{0.2in}
\end{eqnarray}
can be used to express the wave function $\Phi^X({\bf r})$ as
\begin{eqnarray}
\Phi^X({\bf r})=\sum_i\chi_i^X(x,y)\phi_i^X( z).
\end{eqnarray}
If we replace this function in the Schr\"odinger equation (\ref{Hamiltonian}), multiply from the left by $\chi_j^{X*}(x,y)$ and integrate upon $x$ and $y$, we obtain the set of coupled equations
\begin{eqnarray}\label{CouplEq}
-\frac{\hbar^2}{2m}\frac{\partial^2}{\partial z^2}\phi_j^X( z)\!+\!\sum_{i=1}^{N_X}V_{ij}^X(z) \phi_i^X( z)\!=\!(E-\varepsilon_j^X)\phi_j^X( z).\hspace{0.2in}
\end{eqnarray}

Here $N_X$ is the number of propagating modes in layer $X$, or the number of open channels (defined by the condition $E>\varepsilon_j^X$), and
\begin{eqnarray}\label{CouplingMat}
V_{ij}^X(z)\!=\!\!\int_0^{w_x}\!\!\int_0^{w_y}\!dxdy\chi_j^{X*}(x,y) V_{X}^L(x,y,z)\chi_j^X(x,y),\hspace{0.3in}
\end{eqnarray}
are the coupling-channels matrix elements. In this way the 3D multichannel problem is reduced into the 1D multichannel problem. It was shown in Refs. [\onlinecite{PereyraPRL}] and [\onlinecite{PereyraJPA}], and mentioned before, that a general solution for the 1D multichannel periodic system can be obtained in terms of the matrix polynomials $p_n$, when the single-cell transfer matrix $M(z_{i+1},z_{i})$ is known. In actual semiconductor layers, the number of propagating modes depends on the Fermi energy and the cross section $w_xw_y$. When the multichannel problem for a specific semiconductor $X$, with $n_X$ unit cells is solved, one obtains the $N_Xn_X$ energy eigenvalues $E^{X}_{\epsilon,\eta}$ (which determine the conduction and valence bands) and the corresponding eigenfuntions $\phi^{X}_{\epsilon,\eta} (z)$. In the widely used 1D one channel approximation, with $V_{X}(z)=V_{11}^X(z)$, $E^X=E-\varepsilon_1^X$ and $\phi^X(z)=\phi_1^X( z)$, equation (\ref{CouplEq}) becomes
\begin{eqnarray}\label{Hamiltonian1}
\left( \frac{p_z^2}{2m} + V_{X}(z) \right) \phi^X( z)=E^X\phi^X(z).
\end{eqnarray}
\begin{figure*}
\begin{center}
\includegraphics [width=440pt]{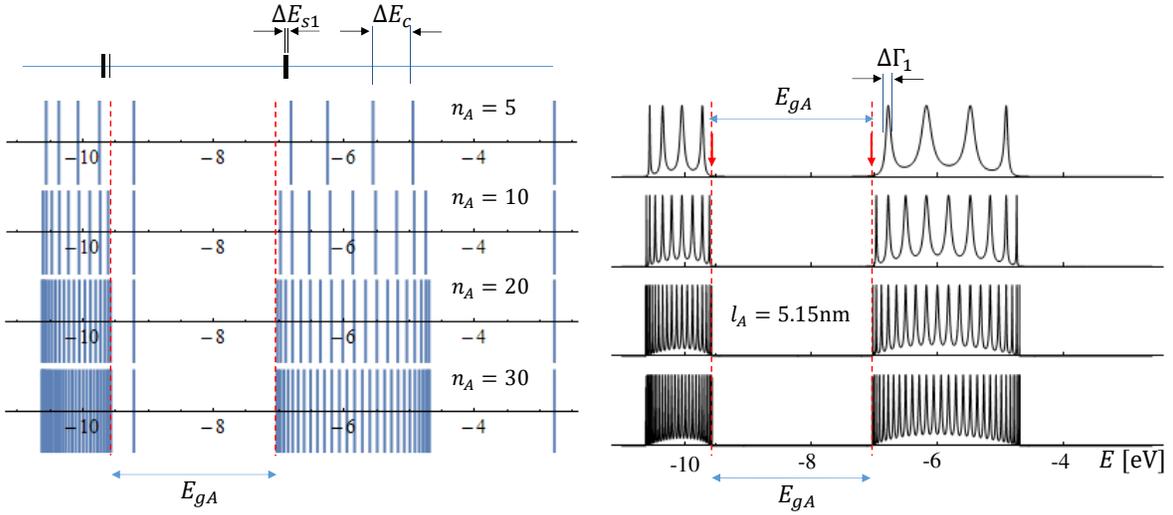}
\caption{Band structure of a bounded periodic semiconductor $C/(a_A)^{n_A}/C$, with $E_{gA}$=2.6eV, $E_{gC}$=3.5eV, and the transmission coefficient through the open SL $(a_A)^{n_A}$. These quantities are plotted  for different layer widths $w_A=l_An_A$. Here $l_A$ is the unit cell length and $n_A$ the number of unit cells in layer $A$ along the growing direction of the heterostructure. On the upper graph of the left hand side, we have also the subbands of the superlattice $C/(AB)^{n}/C$ with $w_B \sim w_A$, $E_{gB}=$2.9eV and $n=$10. }\label{Fig4e}
\end{center}
\end{figure*}

In this limit and given the periodic atomic potentials $V_A(z)$ and $V_B(z)$, in the semiconductor layers $A=(a_A)^{n_A}$ and $B=(b_B)^{n_B}$, one can obtain the unit-cell transfer matrices $M_a$ and $M_b$ and determine, applying the TFPS, the band structures $E^{A}_{\epsilon,\eta}$ and $E^{B}_{\epsilon,\eta}$, and  using the Eq. (\ref{EquEigFunc1}), the eigenfunctions $\phi^{A}_{\epsilon,\eta} (z)$ and $\phi^{B}_{\epsilon,\eta} (z)$. A very good approximation for the atomic potentials $V_A(z)$ and $V_B(z)$, are the effective potentials in the Hartree-Fock approximation. The quantum numbers $\epsilon$ denote the bands, and the quantum numbers $\eta$ the intra-band energy levels.  We will denote the valence and the conduction bands with $\epsilon$=c=1 and $\epsilon$=v=2, respectively.  The intra-band energy levels correspond to $\eta=$1, 2, ... , $n_X$+1. In terms of these energies the fundamental energy gap in layer $X$ is given by
\begin{eqnarray}
 E^X_g\!=\!E^X_{1,1}\!-\!E^X_{2,n_X+1}\!\equiv \!E^X_{{\rm c},1}\!-\!E^X_{{\rm v},n_X+1} \hspace{0.2in}X=A,B.\hspace{0.2in}
\end{eqnarray}
with $E^X_{{\rm c},1}$ the first energy eigenvalue of the conduction band, i.e. the conduction band-edge denoted later as $E^X_{\epsilon_{0X}}$, and $E^X_{{\rm v},n_X+1}$ the last energy eigenvalue of the valence band, i.e. the upper-edge of the valence band. As is well known, the band edges of layers $A$ and $B$ do not coincide, in general (see figure \ref{f2}), and their difference gives rise to the conduction and valence band split offs, as well as, to piecewise constant superlattice or heterostructure potential. \cite{Bastard1987NATO} We will assume from here on that the semiconductor layers $A$ and $B$ are such that $E^A_g < E^B_g $. If energies are below the barrier height ($E<V_b$), see inset in figure \ref{f2}, the eigenfunctions $\phi^{A}_{\epsilon,\kappa}(z)=\phi^{A}(z,E)|_{E=E^{A}_{\epsilon,\eta}}$ are propagating functions while $\phi^{B}(z,E)|_{E=E^{A}_{\epsilon,\eta}}$ are evanescent.\cite{Pereyra2005}

For each value of the quantum number $\eta$ we have the corresponding wave number $\kappa_{\eta}$. To keep some analogy with conventional notation, we can represent the energy eigenvalues $E_{\epsilon,\eta}$ as $E_{\epsilon,\kappa_{\eta}}$ or just as $E_{\epsilon,\kappa}$, that can be written also as $E_{\epsilon}(\kappa)$, keeping in mind that $\kappa$ is discrete.

It is clear that if we are able to determine the eigenvalues $E^{A,B}_{\epsilon,\kappa}$ and eigenfunctions $\phi^{A,B}_{\epsilon,\kappa}$, we are close to obtain the full solution for the heterostructure or SL. Having the wave functions $\phi^{A,B}_{\epsilon,\kappa}$, we must still fulfill the continuity and boundary conditions at the layered structure interfaces. Although this task could, in principle, be accomplished, it is not so simple for these functions (as for the envelope functions) and it is not our purpose here. We will, instead, turn our attention into the derivation of the effective mass approximation based on the existence of the set of rapidly-varying orthogonal functions $\phi^{A,B}_{\epsilon,\kappa}$.

To derive the EMA in the TFPS we need to expand the heterostructure or SL wave functions $\psi(z)$ in terms of the local wave functions $\phi^{A}_{\epsilon,\kappa}(z)$ and $\phi^{B}_{\epsilon,\kappa}(z)$, defined inside the layers $A$ and $B$ respectively. To simplify the discussion let us assume that we have the SL $(AB)^{n}A$. If  $\zeta = z{\rm \, mod}\,[ l_c{\rm]}-a$, with $a$ the width of layer $A$, $l_c=a+b$ the length of the SL unit-cell, and H(w) is the Heaviside function, we can write a rapidly-varying wave function as (see figure \ref{f3})
\begin{eqnarray}\label{SLWaveFunction}
\Phi^{\epsilon_B,\kappa_B}_{\epsilon_A,\kappa_A}(z)&=&{\rm H(}\!-\zeta{\rm )}\phi^A_{\epsilon_A,\kappa_A}(z{\rm \, mod} [l_c{\rm ]})+\cr &&{\rm H(}\zeta{\rm )}\phi^B_{\epsilon_B,\kappa_B}(z{\rm \, mod} [l_c{\rm ]}).
\end{eqnarray}

As mentioned before, in the conventional derivations of the effective-mass approximation, the wave functions inside each layer are expanded in terms of the periodic parts of the band-edge Bloch functions, $u_{l,k_0}^A$ or $u_{l,k_0}^B$, which are generally assumed to be equal.\cite{Enderleinpg252,Bastardpg67} Setting up the SL wave function $\psi(z)$, the assumptions of only one-band and small {\bf k}-vectors are also made.\cite{Enderleinpg252} In the theory of finite periodic systems, the bands and wave functions $\phi^{A}_{\epsilon,\kappa}(z)$ and $\phi^{B}_{\epsilon,\kappa}(z)$ are the energy eigenvalues and the eigenfunctions of the periodic systems $(a_A)^{n_A}$, $(b_B)^{n_B}$. In figure \ref{Fig4e} we show a simplified calculation in the TFPS of the energy spectrum\cite{simplified} and transmission coefficients for a specific (confined and open) semiconductor $A=(a_{A})^{n_A}$, with energy gap $E_{gA}\simeq $2.6eV and unit-cell length $l_A$= 5.15nm.  On the left hand side of figure \ref{Fig4e}, we show the valence and the conduction bands (VB and CB) of the periodic sequence $(a_A)^{n_A}$ bounded by cladding layers $C$, and, on the  right hand side, the transmission coefficients through the same semiconductor but open. At the top of the left hand side column, we plot also the subbands (or minibands) of the SL $(AB)^n$ for $E_{gA}\simeq $2.6eV, $E_{gB}\simeq $2.9eV, $l_A\sim l_B$ 5.15nm and $n$=10. These graphs show that as the layer width  $w_A=l_An_A$ gets thinner, the energy levels separation, $\Delta E_c$, and the  energy-levels widths, $\Gamma E_{\mu}$, increase. On the other hand, it is known that whereas the energy gap $E_{gA}$ remains constant when the number of unit cells $n_A$ varies, the subbands of the superlattice $(AB)^n$, for a fixed barrier width $w_B$, move with the band-edge energy level upwards when $n_A$ decreases, and downwards when $n_A$, hence $w_A$, increases.  This behavior of the energy spectra, justifies the one-band `ansatz' and strengthens the relevance of the band-edge functions as the number of unit cells $n_{A}$ gets smaller. In the specific example of figure \ref{Fig4e}, the level width  $\Delta \Gamma_1 $  is of the order of the subband widths $\sim $ 10meV), and the energy levels separation for a semiconductor with $n_A \sim$5 ($w_A\sim$ 25nm) is approximately 600meV, which is much larger than the bands split off in the conduction and valence bands of layers $A$ and $B$. Thus, in order to define the heterostructure or SL wave function $\psi (z)$ in terms of the envelope and the fast-varying functions, it is justified to consider the band-edge and one-band assumptions. Therefore, we can consider the expansion
\begin{eqnarray}\label{ExpansionFunc}
\psi(z)= \sum_{\kappa^A_{0},\kappa^B_{0}}\langle \epsilon_0,\kappa_0 |\psi \rangle \Phi_{\epsilon_{0},\kappa_{0}}(z).
\end{eqnarray}
Here and in the following,  the quantum numbers $\epsilon_0$ and $\kappa_0$ represent the set $\epsilon^A_{0},\epsilon^B_{0}$ and $\kappa^A_{0},\kappa^B_{0}$, respectively. For a simple and compact notation, we will denote the expansion coefficient $\langle \epsilon_0,\kappa_0 |\psi \rangle$, known also as the envelope function, as $\varphi^{\epsilon_0}_{\kappa_0}(z)$ or $\varphi^{\epsilon_0}(\kappa_0,z)$. If we introduce the  function $\psi(z)$ of Eq. (\ref{ExpansionFunc}) into the SL Schr\"odinger equation
\begin{eqnarray}\label{SLHamiltonian}
\left( \frac{{p_z}^2}{2m} + V_{SL}(z) \right) \psi(z)=E\psi(z),\hspace{0.3in}
\end{eqnarray}
where
\begin{eqnarray}
V_{SL}(z)={\rm H(}\!-\zeta{\rm )}V_A(z\,{\rm mod}  [l_c{\rm ]})+{\rm H(}\zeta{\rm )}V_B(z \,{\rm mod} [l_c{\rm ]}), \hspace{0.3in}
\end{eqnarray}
multiply by $\Phi_{\epsilon_{0},\kappa'_{0}}(z)$ and integrate, we have
\begin{eqnarray}
\sum_{\kappa^A_0,\kappa^B_0}\!\!\left[{\rm H(}\!-\!\zeta{\rm )}E_{\epsilon^A_{0},\kappa^A_0}\delta_{\kappa^A_0,\kappa^{A'}_{0}}\!\! + \!{\rm H(}\zeta{\rm )}E_{\epsilon^B_{0},\kappa^B_0}\delta_{\kappa^B_0,\kappa^{B'}_{0}} \right]\langle \epsilon_0,\kappa_0 |\psi \rangle\!\!\cr =\! E \langle \epsilon_0,\kappa_0 |\psi \rangle.\hspace{0.5in}
\end{eqnarray}
Since
\begin{eqnarray}
\!E_{\epsilon^B_{0},\kappa^B_{0}}\!=\!E_{\epsilon^A_{0},\kappa^A_0}\!+
\!V_{\kappa^A_{0},\kappa^B_{0}}\!=\! E_{\epsilon^A_{0},\kappa^A_{0}}\!+\!\langle \kappa^A_{0}|V_{\epsilon P}|\kappa^B_{0} \rangle , \hspace{0.3in}
\end{eqnarray}
the sectionally constant periodic potential $V_P(z)$, known as the split off, {\it appears here naturally as a consequence of the difference in the energy band structures of  layers $A$ and $B$}, both in the conduction and valence bands. Therefore, we are left with
\begin{eqnarray}
E^{A}_{\epsilon^A_{0}}(\kappa^{A}_0)\varphi^{\epsilon_0}(\kappa^{A}_{0}) \!+\! \sum_{\kappa^B_{0}}\langle \kappa^{A}_{0}|V_{\epsilon P}|\kappa^B_{0} \rangle \varphi^{\epsilon_0}(\kappa^B_{0}) \!=\! E \varphi^{\epsilon_0}(\kappa^{A}_{0}).\nonumber \\
\end{eqnarray}
We can now, as usual, multiply by $(1/\Omega) e^{i \kappa z}$  and sum the Fourier series to obtain
\begin{eqnarray}
E^A_{\epsilon_0}(-i\frac{\partial}{\partial z}){\it \Psi}^{\epsilon_0}(z) +V_P (z) {\it \Psi}^{\epsilon_0}(z) = E {\it \Psi}^{\epsilon_0}(z).\hspace{0.3in}
\end{eqnarray}
If we further approximate $E_{\epsilon_0}(-i\partial/\partial z)$ by a quadratic function of $-i\partial/\partial z$, near the band edge, assuming that the k-vector at the edge is small and an effective mass $m^{*}_{\epsilon_0}$, defined as usual for each layer, we have
\begin{eqnarray}\label{effectivemasseq}
\left[\frac{p_z^2}{2 m^{*}_{\epsilon_0}}  +V_P (z)\right] {\it \Psi}^{\epsilon_0}_{\mu,\nu}(z) = (E -E^A_{\epsilon_0,\eta_0})_{\mu,\nu}{\it \Psi}^{\epsilon_0}_{\mu,\nu}(z),\hspace{0.3in}
\end{eqnarray}
with $\epsilon_0=c$ and $\eta_0$=1 for the conduction band and $\epsilon_0=v$ and $\eta_0$=$n_A$+1 for the valence band. If we define the energy eigenvalues
\begin{eqnarray}
 E_{\mu,\nu}=(E -E^A_{\epsilon_0,\eta_0})_{\mu,\nu},
\end{eqnarray}
measured from the band edges, we can write the Schr\"odinger equation in the effective mass approximation
\begin{eqnarray}\label{effectivemasseq}
\left[\frac{p_z^2}{2 m^{*}_{\epsilon_0}}  +V_P (z)\right] {\it \Psi}^{\epsilon_0}_{\mu,\nu}(z) = E_{\mu,\nu}{\it \Psi}^{\epsilon_0}_{\mu,\nu}(z),\hspace{0.3in}
\end{eqnarray}
that we were looking for and was used for SLs and heterostures, without a specific proof. As mentioned before, for SLs we can use the TFPS to solve this equation and to determine the eigenvalues $E_{\mu,\nu}$ and the eigenfunctions ${\it \Psi}^{\epsilon_0}_{\mu,\nu}(z)$, known as envelope functions. It is worth noting that this derivation of EMA does not require that the layered structure be periodic. Therefore, the EMA is valid for any layered heterostructure.

\begin{figure}
\begin{center}
\includegraphics [width=240pt]{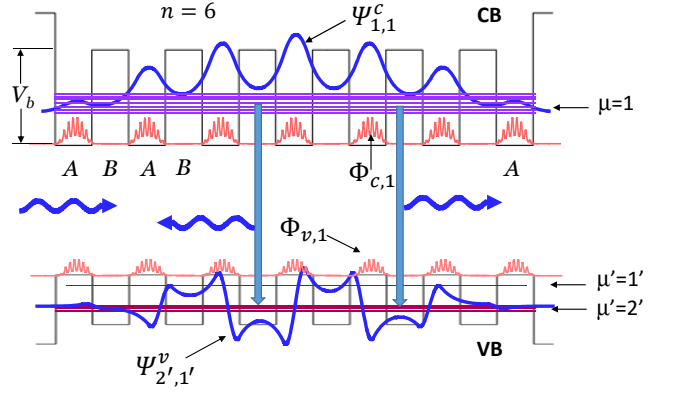}
\caption{Edge and envelope functions in the conduction and valence bands. The edge functions $\Phi_{2,1}(z)\equiv \Phi_{c,1}(z)$ and $\Phi_{1,1}(z)\equiv \Phi_{v,1}(z)$ decay exponentially in the SL barriers. The SL eigenfunctions ${\it \Psi}^{c}_{\mu,\nu}(z)$ and  ${\it \Psi}^{v}_{\mu',\nu'}(z)$ in the conduction and valence bands are the extended envelope functions, relevant in the optical response calculations. We show here the functions ${\it \Psi}^{c}_{1,1}(z)$ and ${\it \Psi}^{v}_{2',1'}(z)$ in the subbands $\mu$=1 and $\mu'$=$2'$.} \label{f5}
\end{center}
\end{figure}
All the assumptions behind this derivation imply that the wave functions $\psi(z)$ can be written as
\begin{eqnarray}\label{WaveFunction}
\psi(z)\rightarrow {\it \Psi}^{\epsilon_0}_{\mu,\nu}(z)  \Phi_{\epsilon_0,\eta_0}(z)
\end{eqnarray}
with ${\it \Psi}^{\epsilon_0}_{\mu,\nu}(z)$ the SL eigenfunction (envelope functions) and  $\Phi_{\epsilon_0,\eta_0}(z)$ the rapid oscillating wave functions. In figure \ref{f5} we plot the functions ${\it \Psi}^{c}_{1,1}(z)$ and $ \Phi_{c,1}(z)$, in the conduction band, and the functions ${\it \Psi}^{v}_{2',1'}(z)$ and $\Phi_{v,1}(z)$ of the valence band. these functions can in principle be determined within the TFPS.

Dealing with transport properties, one can neglect the  function $\Phi_{\epsilon_0,\eta_0}(z)$, however, for calculations involving two bands, the whole wave function $\psi (z)$ should, in principle, be considered. We will show now that the fast-varying factor $\Phi_{\epsilon_0,\eta_0}(z)$ can effectively be ignored in optical response calculations.

\section{On the redundancy of the fast-varying functions}

\begin{figure}
\begin{center}
\includegraphics[width=240pt]{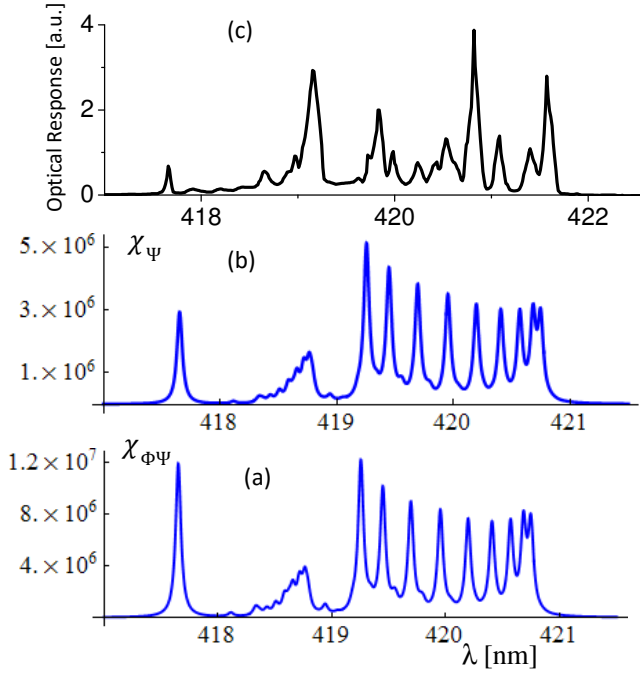}
\caption{The effect of the fast-oscillating functions on the optical response. In panels (a) and (b) the calculated optical response to describe the experimental spectrum in panel (c) for the blue-emitting $GaN/InGaN$ superlattice. In panel (a) the optical response
$\chi_{_{\Phi{\it \Psi}}}$ is calculated by taking into account fast-oscillating wave functions $\Phi_{\epsilon_0,\eta_0}(z)$. In panel (b) the optical response
$\chi_{_{{\it \Psi}}}$ is calculated by ignoring rapidly-oscillating wave functions. The experimental spectrum is reproduced with permission from [\onlinecite{NakamuraPaper}]. Copyright [1996], AIP Publishing LLC.}\label{fig 6}
\end{center}
\end{figure}

To determine the effect of the rapidly-oscillating factor $ \Phi_{\epsilon_0,\eta_0}(z)$ on the optical response, let us consider the blue emitting $(In_{0.2}Ga_{0.8}N\backslash In_{0.05}Ga_{0.95}N)^{10}\backslash In_{0.2}Ga_{0.8}N$ superlattice studied in Refs. [\onlinecite{NakamuraPaper}] and [\onlinecite{PereyraEPL}]. We will calculate the optical response
\begin{eqnarray}\label{susceptRF}
\text {\font\tenxii=cmr6
\scriptscriptfont0=\tenxii
$\chi_{_{\Phi{\it \Psi}}}$ }= \sum_{\nu,\nu'}f_{eh}\frac{\displaystyle \Bigl{|}\langle\psi^v_{\rm f} |H_{\rm int}|\psi^c_{\rm i}\rangle\Bigr{|}^{2}}{(\hbar \omega-E_{1,\nu}^{c}+E_{2',\nu'}^{v}+E_B)^{2}+\Gamma^{2}}\hspace{0.2in}
\end{eqnarray}
taking into account the fast-varying functions $\Phi_{\epsilon_0,\eta_0}(z)$, which means $\psi^c_{\rm i}$ = ${\it \psi}^{c,1}_{1,\nu}(z)$ = $\Phi_{c,1}(z){\it \Psi}^c_{1,\nu}(z)$ and   $\psi^v_{\rm f}$ = ${\it \psi}^{v,n_A+1}_{2',\nu'}(z)$ = $\Phi_{v,n_A+1}(z){\it \Psi}^v_{2',\nu'}(z)$. These results are compared in figure (\ref{fig 6}) with the optical response
\begin{eqnarray}\label{susceptNRF}
\text {\font\tenrm=cmr12
\scriptscriptfont12=\tenrm
$\chi_{_{{\it \Psi}}}$ }= \sum_{\nu,\nu'}f_{eh}\frac{\displaystyle \Bigl{|}\langle{\it \Psi}^v_{2',\nu'} |H_{\rm int}|{\it \Psi}^c_{1,\nu}\rangle\Bigr{|}^{2}}{(\hbar \omega-E_{1,\nu}^{c}+E_{2',\nu'}^{v}+E_B)^{2}+\Gamma^{2}}\hspace{0.2in}
\end{eqnarray}
calculated in Ref. [\onlinecite{PereyraEPL}], ignoring the fast-varying functions. As was shown in this reference and can be seen in figure \ref{fig 6}, this optical response agrees extremely well with the experimental results in panel (c). \cite{NakamuraPaper,PereyraEPL} In (\ref{susceptRF}) and (\ref{susceptNRF}), $\hbar \omega$ is the emitted photon energy,  $E^c_{1,\nu}$ the energy levels in the first subband of the CB, $E^v_{2',\nu'}$ the (heavy hole) energy levels in the second subband of the VB, $E_B$ the exciton binding energy, $f_{eh}$ the occupation probabilities and $\Gamma$ the level broadening energy.
\begin{figure}
\begin{center}
\includegraphics[width=240pt]{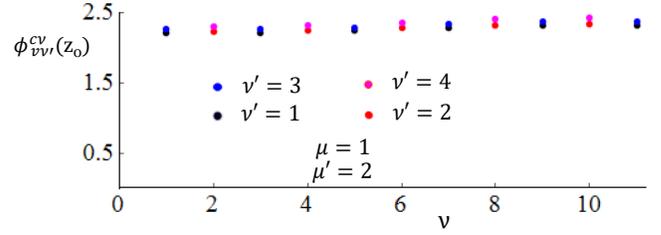}
\caption{The mean value factor $\phi^{c,\nu}_{v,\nu'}(z_0)$ for transition from levels $(\mu,\nu)=(1,1),(1,2),...,(1,11)$, in the first subband of the CB, to $(\mu'\nu')=(2',1'),(2',2'),(2',3')$ and $(2',4'$, of the second subband of the VB.}\label{fig 7}
\end{center}
\end{figure}

Besides the overall amplification, by a factor of $\simeq$ 2.4, our calculations show that the rapidly-varying functions have no effect on the optical spectrum.

According with the mean value theorem for definite integrals, the optical response \text {\font\tenxii=cmr6
\scriptscriptfont0=\tenxii
$\chi_{_{\Phi{\it \Psi}}}$} in equation (\ref{susceptRF}) can be written as
\begin{eqnarray}\label{NsusceptPL}
\text{\font\tenxii=cmr6
\scriptscriptfont0=\tenxii
$\chi_{_{\Phi{\it \Psi}}}$ }= \sum_{\nu,\nu'}\phi^{c,\nu}_{v,\nu'}(z_0)\text {\font\tenrm=cmr12
\scriptscriptfont12=\tenrm
$\chi_{_{{\it \Psi}}}$ }
\end{eqnarray}
with $\phi^{c,\nu}_{v,\nu'}(z_0)$ a number, which in principle depends on the quantum numbers $\nu$ and $\nu'$. Specific calculations show that this factor is almost constant (see figure \ref{fig 7}), and consistent with the differences in the numerical values of the optical responses $\chi_{_{\Phi{\it \Psi}}}$ and $\chi_{_{{\it \Psi}}}$ in figure \ref{fig 6}.

\section{Conclusions}

We have derived the effective mass approximation for the Schr\"odiger equation of layered hetrostructures, based on the energy eigenvalues and rapidly- oscillating eigenfunctions obtained, for each layer, in the theory of finite periodic systems. This derivation that is based on physical quantities of finite structures explains why the EMA works so well when applied to this kind of systems. We have shown also that, in order to calculate interband transition matrix elements, the rapidly-oscillating wave functions $\Phi_{\epsilon_0,\eta_0}(z)$, that should be multiplied by the envelope functions, ${\it \Psi}^{\epsilon_0}_{\mu,\nu}(z)$, can safely be ignored.

\section{Acknowledgement}

I acknowledge the useful comments of Herbert P. Simanjuntak.


\begin{references}
\bibitem{simplified} The spectrum and transmission coefficients in this figure were obtained for a simplified and approximate piecewise constant potential. The potential parameters were chosen to fit the accurate results that will be published elsewhere.
\end{references}
\end{document}